\documentclass[pre,amsmath, twocolumn, floatfix,showpacs]{revtex4}
\usepackage{graphicx}
\usepackage{amssymb}
\usepackage{amsmath}
\newcommand{\lo}{\langle} \newcommand{\rc}{\rangle}

\begin{document}

\title{efficient path sampling on  multiple reaction channels}
\author{Titus S. van Erp} \affiliation{Centrum voor Oppervlaktechemie
en Katalyse, K.U. Leuven, Kasteelpark Arenberg 23, B-3001 Leuven,
Belgium}

\begin{abstract}
Due to the time scale problem, rare events are not accessible  by straight forward molecular dynamics.  The presence of  multiple  
reaction channels complicates the problem even further. The feasibility of the standard 
free energy based 
methods
relies  strongly on the success in finding a proper  reaction coordinate. This can be 
very difficult task in  high-dimensional complex systems and even more if several  distinct 
reaction channels exist. Moreover, even if a proper reaction coordinate can be found, ergodic sampling will be a challenge. In this article, we discuss the recent advancements of path sampling methods
to tackle this problem. We argue why the path sampling methods, via the transition interface sampling  technique, is less sensitive to the choice of 
reaction coordinate. Moreover, we review a new algorithm, parallel path swapping, that can dramatically improve the ergodic sampling of trajectories for
the multiple reaction channel systems.

\end{abstract}
\maketitle

\section{Introduction}
Path sampling has shown to be efficient tool to study rare events
that are not accessible by straight forward molecular dynamics (MD).
The principal idea behind the method~\cite{TPS98} is to reduce the superfluent exploration of the
stable states and to generate trajectories that have a high chance to 
become reactive. Via the transition interface sampling (TIS) 
method~\cite{ErpMoBol2003,MoErpBol2004,titusthesis,ErpBol2004},
sets of trajectories can be generated that progressively
climb across the barrier towards the product state.  
Besides the rate of the reaction, the TIS path ensembles allow  to analyze the 
mechanism. Each TIS simulation in the series give a correct distribution
of pathways that reach a certain level towards reactivity.  Some of them will 
fail to reach a next level and some  of them will successfully
make a step further. 
Therefore, these path ensembles yield a treasure of information for 
analyzing  complex reaction mechanisms in a very intuitive way.
The main TIS equation relates the reaction rate with a flux through a surface closeby the
reactant state times the overall crossing probability
\begin{align}
\label{kTIS}
k_{AB} &=f_A  {\mathcal P}_A(\lambda_B|\lambda_A).
  \end{align}
  The flux $f_A$ is simply the number of crossings with surface $\lambda_A$ per unit time. 
${\mathcal P}_A(\lambda_B|\lambda_A)$
is the probability that whenever the surface $\lambda_A$ is crossed,
the system will go all the way over the barrier to cross the surface 
$\lambda_B$ at the other side. 
The surfaces are generally defined by a parameterization using a single parameter called
reaction coordinate (RC) or order parameter. The RC can be any non-linear function
of all particle positions and velocities in the system~\footnote{Normally,  configuration space will be sufficient. However, to ensure the stability of states $A$ and $B$, the surfaces $\lambda_A$ and $\lambda_B$ might require explicit kinetic energy dependence (see e.g.~\cite{ErpMoBol2003}). }. The surface (or interface) $\lambda$
is then simply defined as the collection of phase points for which this function is exactly
$\lambda$. Typical examples of RCs are  distances of bonds that have to broken or formed,
the size of the largest solid cluster in nucleation studies, or the number of native bonds
for protein folding. By convention, we assume that $\lambda_A < \lambda_B$.
It is important to realize that the reaction rate $k$ does not depend on the choice of the function RC or on the values
$\lambda_A$ and $\lambda_B$ as long as they obey some very basic
principles. That is, each trajectory from $\lambda_A$ to $\lambda_B$ should be a true
reactive event of the reaction of interest.  In addition, the chance, that  the system quickly returns
to $\lambda_A$ after $\lambda_B$ is crossed, should be of the same order as that of
an independent reverse reaction.

$f_A$ can be computed by
straight-forward MD. 
${\mathcal P}_A(\lambda_B|\lambda_A)$ is very small and therefore difficult to compute.
However, if we define a set of interfaces in between with $\lambda_0=\lambda_A, \lambda_n=\lambda_B$, and $\lambda_i < \lambda_{i-1}$,  we can invoke following exact 
factorization: 
\begin{align}
\label{PTIS}
 {\mathcal P}_A(\lambda_B|\lambda_A) &=  {\mathcal P}_A(\lambda_n|\lambda_0)
=\prod_{i=0}^{n-1} {\mathcal P}_A(\lambda_{i+1}|\lambda_i).
  \end{align}
Here, ${\mathcal P}_A(\lambda_{i+1}|\lambda_i)$ is a history dependent conditional crossing probability that
is much higher than ${\mathcal P}_A(\lambda_B|\lambda_A)$ and, therefore,  much easier to compute.
It equals the probability that, given the system is about to cross $\lambda_i$ for the first time
since its last crossing with $\lambda_A$, it will cross $\lambda_{i+1}$ as well before $\lambda_A$.
Computing this factor can be done by generating a representative set of trajectories that start at $\lambda_A$,
cross $\lambda_i$, and end at either $\lambda_A$ or $\lambda_{i+1}$. The fraction that ends at  $\lambda_{i+1}$ equals
${\mathcal P}_A(\lambda_{i+1}|\lambda_i)$. Recent TIS simulations employ a somewhat different ensemble.
They consider all possible pathways that start at $\lambda_A$, end at either  $\lambda_A$ or $\lambda_B$, and
have at least one crossing with $\lambda_{i}$. ${\mathcal P}_A(\lambda_{i+1}|\lambda_i)$ is then the fraction
that crosses $\lambda_{i+1}$. Generating this set is only slightly more expensive as
most trajectories will end at $\lambda_A$ anyway. However, this approach makes it much easier 
to initialize the TIS parameters, such as the  interface positions, while running one path simulation after the other in the simulation series. Moreover, the implementation of  the new path swapping technique is 
much more straightforward  when these ensembles are used.  
Of course, the crucial point is how to generate these trajectories. In principal we could 
simply run a MD trajectory and cut out the pieces of interest, but then our efficiency would be as bad as MD.
Luckily, Dellago et al. developed efficient Monte Carlo moves 
to generate these trajectories
within the context of transition path sampling (TPS)~\cite{TPS98_2}.  The algorithm only requires  to provide a single trajectory that satisfies the conditions. 
The principal MC move, called shooting, picks a random
timeslice of the old trajectory, changes the velocities  by a small amount, and integrates
the equations of motion 
 backward and forward in time to obtain a new trajectory.  This trajectory can be accepted if it also 
satisfies the opposed  conditions. The MC move in TIS is a slight adaptation of this shooting move
to allow flexible path lengths (for a graphical illustration of the TIS shooting algorithm see \cite{vanErp07}).  This flexible path length is a strong improvement upon the original TPS
algorithm. The trajectories in TIS will on average be shorter than in TPS. 
In addition, TIS does not introduce
any systematic error as the small fraction of long trajectories is also taken into account. These will usually lie outside the range of path lengths one would consider in TPS. Other improvements are that the TIS
formulation of the reaction rate is less sensitive to recrossings and that TPS shifting moves have become
redundant in the TIS algorithm.   

The TIS algorithm has been successfully applied to several systems ranging from protein folding~\cite{bolhuisPNAS},
nucleation~\cite{Moroni05},  micelle fusion and fission~\cite{pool07}, 
and DNA denaturation~\cite{vanErp07PRL}. Moreover, the TIS theory has initiated the development 
of several new path sampling methods such as the partial path TIS (PPTIS)~\cite{MoBolErp2004} and forward flux sampling (FFS)~\cite{FFS}.  PPTIS was especially designed for diffusive barrier crossings. It uses a history-loss assumption to reduce the path length even further. 
FFS is exact, like TIS, but uses another type of MC move in which one 
integrates the equations of motion only forward in time . 
The advantages of FFS is that it works for both equilibrium and out of 
equilibrium systems and that the dynamics do not necessarily  have to be 
reversible. The disadvantage is that FFS introduces much stronger correlations 
between data points.
This has a negative effect on the efficiency of the method, which can become 
extreme if the dynamics bear a strong deterministic character or when the RC 
is not well chosen~\cite{van06}.  Both PPTIS and FFS are based on Eqs.~(\ref{kTIS}) and (\ref{PTIS}).

In this article we will discuss the computational challenging  problem of  multiple reaction channels.
This problem is quite generic for any transition that involves many important degrees of freedom
and it poses several difficulties. First of all it is extremely difficult to derive good RCs (like committor
surfaces~\cite{Bolhuis02,luca06})
 for such a system
that can be efficiently be used in a free energy based approach. Therefore, one would either need 
a systematic  approach to obtain feasible RCs or a method that is much less sensitive to the choice of RC than the standard methods. Secondly, in order to explore one channel after the other, one needs
a MC sampling that is highly nonlocal. Finally, the method should be very flexible and 
able to let the nonlocal
MC moves coincide with a high rate of acceptance.
We will show that TIS encompasses  both the required insensitivity to the RC and the nonlocal character
of shooting moves. Moreover, using  the  additional new  technique of parallel path swapping (PPS),
we can also fulfill the last requirement upto a very satisfactorily level.  This results in a very efficient approach to treat  multiple reaction channel systems. This article is organized as follows. In Sec.~\ref{secRC}, we discuss the issue of RC sensitivity for TIS and related methods. 
In Sec~\ref{secPS} we introduce the new PPS technique and
review recent results, based on this method,  for the DNA denaturation~\cite{vanErp07PRL}.
We end with conclusions in Sec.~\ref{seccon}.

\section{Reaction Coordinate dependence}\label{secRC}
The problem of finding suitable RCs in high-dimensional complex systems has been an outstanding problem for several decades. The free energy based approaches, pioneered by 
Wigner~\cite{W38},  Eyring\cite{E35}, and  Keck \cite{Keck62}  and further developed by  Bennett~\cite{Bennet77}, Chandler~\cite{DC78},  rely on a two-step approach. First the free energy as function of the 
proposed RC needs to be calculated using umbrella sampling (US)~\cite{TV74}  or 
thermodynamic integration (TI)~\cite{CCH89} techniques. From this, the free energy barrier can be derived and the transition state theory (TST) approximation for the rate.  For complex systems, the TST approximation needs to be corrected for fast recrossing events by a dynamical factor $\kappa$ (with $0 < \kappa \leq 1$), called the transmission coefficient. If $\kappa$ is not too low, it can be accurately determined by releasing dynamical trajectories forward and backward in time from the top of the free energy barrier. $\kappa$ is then related  to the fraction of trajectories that connect reactant state $A$ with product state $B$~\footnote{It is a bit more complicated than this. In addition, one needs to avoid the overcounting of
successful trajectories that cross the top of the barrier multiple times and each trajectory needs to be weighted with its initial velocity at the top along the direction of the RC~\cite{ErpBol2004}.}.
In this way, one can determine $k$ exactly, independently to the choice of RC, 
and one obtains
a set of reactive trajectories that can be subject of further investigations.
 However, if the RC fails to capture 
the actual mechanism, the vast majority of trajectories are either of the type $A \rightarrow A$ or $B \rightarrow B$ and $\kappa$ becomes immeasurably small. Illustrative for this effect are chemical reactions in solution which require considerable solvent rearrangements.
The top of the free energy barrier, defined by a RC that does not include 
the solvent degrees of freedom, will either correspond to the situation where the solvent can easily incorporate the reactant or the product species.  As result, both forward and backward dynamics will rapidly collapse to the same state. In addition,  the evaluation  of the free energy barrier 
itself becomes problematic. Considerable hysteresis will occur in  US or TI 
when the system is dragged over the barrier 
from reactant to product state and back. 

This RC problem was the main driving force to develop alternative methods like TPS~\cite{TPS98}.  Although it is quite evident that TPS  can generate 
reactive trajectories quite  efficiently using  a simple type of RC, 
it is not so clear if the same is true for the actual reaction rate evaluation. 
Only recently, Ref.~[\onlinecite{van06}] really proved, 
by analytical results of a simple  2D model, 
that the TIS efficiency 
of the reaction rate calculation
is much less sensitive to an improper choice of the RC than free energy based methods.  There are two reasons for this. First of all, 
as paths are global identities (the weight of a path does not depend on a single point),
the sampling of paths instead of phase points can eliminate  hysteresis effects~[\onlinecite{van06}]. 
Secondly, whereas free energy based methods apply important sampling approaches to configuration space only, hoping that the dynamical factor will be not too small, path sampling applies importance sampling techniques on the dynamical factor directly. The RC insensitivity  is not generic for any type of
path sampling method. It is  valid for the original TPS method as well, but not for e.g. PPTIS or FFS~\cite{van06}. 
We can also give a third reason which was not important for the 2D example studied in~\cite{van06}.
As the TIS/TPS shooting move is nonlocal, it should be able to circumvent barriers that are 
orthogonal to the RC.  Indeed, a TPS water
trimer study~\cite{Geissler99}  revealed that the shooting algorithm was capable of finding two reaction
mechanisms across different saddle points separated by a barrier higher than
the total energy of the NVE simulation. 
An approach to enhance this effect is explained in the next section.

\begin{figure}[t]
\begin{center}
\includegraphics[width=9cm,keepaspectratio]{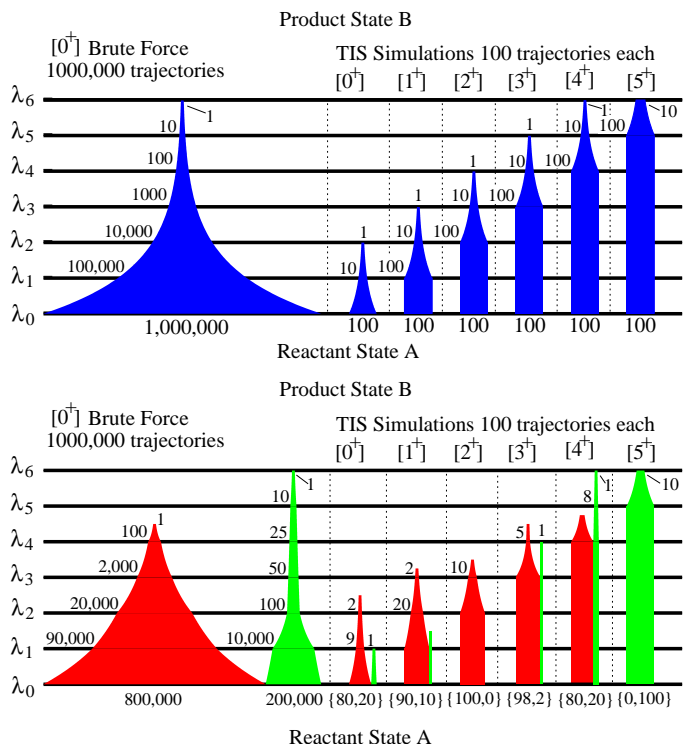}
\end{center}
\caption{(color online) Path survival graphs. Top: a possible case for a one channel system. Interfaces are positioned such that one out of ten reaches the next interface. The left-hand side is the situation when trajectories are generated by brute force using one million trajectories. The right-hand side shows the results  (assuming a perfect sampling) for the TIS simulation series, each simulation consisting of only one hundred trajectories. Bottom:  a possible case for a two channel system. Again a brute force simulation of one million trajectories is compared  
with a TIS simulation series of one hundred trajectories each.
One channel (red) has initially a higher survival rate than the green one, but turns into a dead end. 
The paths gathered by TIS for the $[2^+]$ ensemble do not contain a single green path. 
Still, the product of the obtained
crossing probabilities gives the right result.} 
\label{graph}
\end{figure}
The different dependence on the RC of TIS and FFS, although based on the same equations, can  be explained by the argument that Eq.~(\ref{PTIS}) has two possible
interpretations.  The conditional crossing probabilities can be viewed as a kind of physical 
(non-Markovian) hopping probabilities to go from one interface to the other like climbing up a ladder.
This interpretation is very close to the FFS implementation of Eq.~(\ref{PTIS}) . The alternative interpretation is that  Eq.~(\ref{PTIS})  simply corresponds to a multiplication of ratios between
the number of paths contained by different sets: ${\mathcal P}_A(\lambda_B|\lambda_A)=
\frac{\# \textrm{paths } \in [n^+]}{\# {\rm paths} \in [0^+]} =
\frac{\# \textrm{paths } \in [1^+]}{\# {\rm paths} \in [0^+]} \times
\frac{\# \textrm{paths } \in [2^+]}{\# {\rm paths} \in [1^+]} \times 
\frac{\# \textrm{paths } \in [3^+]}{\# {\rm paths} \in [2^+]} \times \ldots \times
\frac{\# \textrm{paths } \in [n^+]}{\# {\rm paths} \in [(n-1)^+]}
$,
where $[i^+]$ defines the collection of paths that start at $\lambda_A$ and have at least one crossing
with $\lambda_i$ before revisiting   $\lambda_A$ or ending at $\lambda_B$. 
From this, it is directly clear that if we replace
e.g. $[3^+]$ in this factorization by an arbitrary different set of trajectories, this still does not change the validity of the equation. This has an important implication. The final trajectory set  $[n^+]$ can be
fundamentally different than $[3^+]$. They do not necessarily
have to resemble up to $\lambda_3$.  

In Fig.~\ref{graph} we give two simple examples for seven interfaces $\{\lambda_0, \ldots, \lambda_6\}$. Shown here 
are path survival diagrams. 
In Fig.~\ref{graph}-(top), we assume that only 1 out of the $10^6$ trajectories that initially start at $\lambda_A$
will reach $\lambda_B$ and  at each  interface only 10\% survives reaching the next one.   
Hence, if we would simply run straightforward MD trajectories, we would need at least
  $10^6$ trajectories to hope for  a single reactive event.  Now, if we only run 100 trajectories per TIS (or FFS) ensemble, we expect that each time 10 trajectories will reach the next level. Our final result
  ${\mathcal P}_A(\lambda_A|\lambda_B)=\prod_{i=1}^n {\mathcal P}_A(\lambda_i|\lambda_{i-1})=
  (10/100)^6=1\cdot 10^{-6}$ is exactly identical to a perfect brute force calculation but using only
  600 trajectories instead of  $10^6$   and having 10 fully reactive trajectories instead of only 1
  in the end.
  This directly shows the orders of magnitude improvement of the path sampling simulations compared to MD. In a bit more elaborate calculation, one can show that the effective computational cost scales exponentially with the barrier height in brute force  MD and only quadratically in TIS. This is similar to US using rectangular bias windows~\cite{van06}.
In Fig.~\ref{graph}-(bottom), we show a bit more complicated example with two channels. One is initially favorable but turns into a dead end, while the other one survives. The right-hand side shows the most likely 
outcome for the TIS simulation series. It is important to note that at some point, at the $[2^+]$ simulation,
the set of trajectories generated by TIS does not contain a single green path. Still the final
result 
$
{\mathcal P}_A(\lambda_6|\lambda_0)=
(10/100) \cdot (20/100) \cdot(10/100) \cdot
   (6/100) \cdot (8/100) \cdot (10/100) 
=  0.96 \cdot 10^{-6}
$ is correct. This is a very important point to realize as it shows once more the flexibility of the TIS method. FFS which propagates trajectories only forward in time will not be able to recapture the 
green trajectories once they are lost. Hence FFS requires to sample much more trajectories at
the initial interface ensembles or to device a RC where this problem will not occur. 
Of course, in order to recapture the successful green pathway it is of eminent importance that
the TIS shooting move will be able to make this transition. In the next section we show how 
PPS can be a very effective tool to enhance these transitions.
\section{Parallel Path Swapping} \label{secPS}
\begin{figure}[t]
\begin{center}
\includegraphics[width=8cm,angle=0,keepaspectratio]{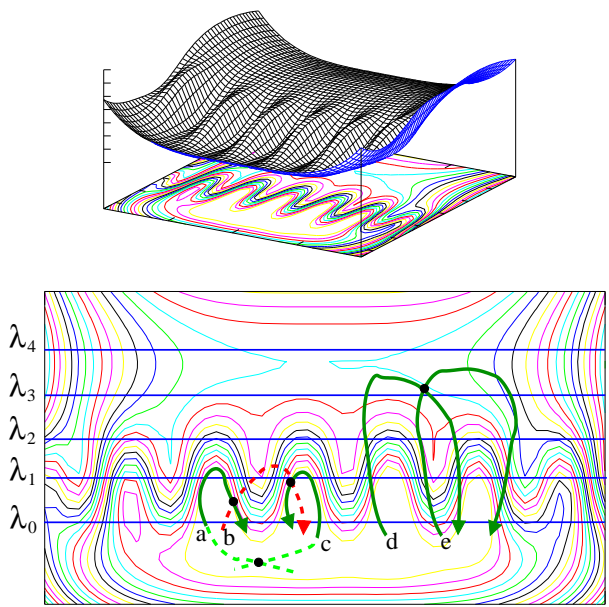}
\end{center}
\caption{(color online) Illustration of the multiple channel barrier. Contour plot is shown below.  Trajectories
{\bf a} and {\bf c} are two possible trajectories in the $[1^+]$ ensemble situated in two different channels.
In order to find trajectory {\bf c} from trajectory {\bf a}, we would either need to shoot (with some luck) from inside the 
reactant well, or to generate the bridging trajectory {\bf b} that is high in energy and thus likely rejected. 
Trajectories {\bf d} and {\bf e} are trajectories in the $[3^+]$ ensemble. They are much higher in energy
and can easily move from one channel to the other.}
\label{channels}
\end{figure}

In Fig.~\ref{channels} we depict the problems that can occur when the standard shooting move is applied for a multiple reaction channel system. 
The trajectories {\bf a} and {\bf c} of the $[1^+]$ ensemble are situated in different channels
of the potential energy barrier. It is not very likely that the shooting move will  be able to connect these trajectories. It would require to generate a bridging trajectory {\bf b}, but such a trajectory is very high in energy and will likely be rejected.  Alternatively, in order to make use of
the non-locality of the shooting move and its ability to avoid transversal barriers, we might want to 
shoot from
inside the reactant well.  However, such a  point is not part of the paths of this ensemble that start
at $\lambda_A$. We might shift the $\lambda_A$ interface more into the reactant well with the expense that all trajectories become longer and thus more expensive. However, even then we must be very lucky
because if we fail to cross $\lambda_1$, this trajectory must be rejected and we remain in the same channel. On the other hand, if we consider the trajectories of the $[3^+]$ ensemble, these are much higher in energy and can easily move from one channel to the other.  
It would be very useful if we could somehow make advantage of the 
high energy paths of the $[3^+]$ ensemble and the non-local shooting moves 
from inside the reactant well.

A very successful method in standard MC is the parallel tempering or replica exchange method~\cite{marinari92}. In this  method  one performs several simulations in parallel at different temperatures. Then, with a certain frequency and acceptance probability the configurations at one temperature simulation is being swapped with a lower temperature simulation. The high temperatures will easily explore the rough free energy surface as they have a much lower probability to get trapped.
The low temperature simulations will benefit from the exchange of information as they will be able to hop from one potential basin to another 
without  having to cross the intermediate barriers physically.
The combination of path sampling and parallel tempering has been used before~\cite{VS01}, but the main disadvantage of this approach is that one needs to perform expensive additional simulations
whose information is useless if one is interested in the reaction rate at one temperature alone. In addition,
parallel tempering will not help to circumvent entropic barriers using the non-locality of the shooting move. 
However, if one realizes that the TIS method already consists of a series of simulations,
it makes sense to introduce a swapping move between these~\cite{vanErp07PRL}. 

\begin{figure}[t]
\begin{center}
\includegraphics[width=5cm,angle=0,keepaspectratio]{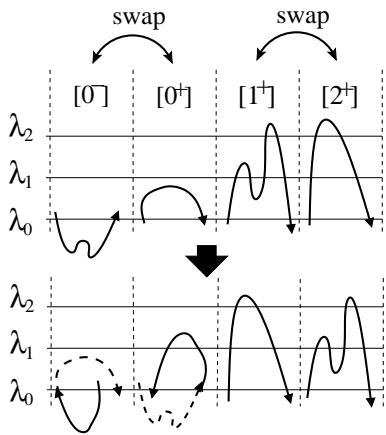}
\end{center}
\caption{Illustration of the swapping moves $[0^-] \leftrightarrow [0^+]$ and 
$[1^+] \leftrightarrow [2^+]$. The alternative swapping possibility  
$[0^+] \leftrightarrow [1^+]$
would have yielded a rejection as the $[0^+]$ path does not cross $\lambda_1$ and is therefore not a valid 
path for the $[1^+]$ ensemble.}
\label{swap}
\end{figure}

In order to create the maximum flexibility of swapping moves at all levels, it is convenient to replace the
initial MD simulation for the flux calculation $f_A$ by another type of path ensemble $[0^-]$.
These consists of all the paths that start   at
$\lambda_A$, then go initially towards the negative direction, 
and finally end again at $\lambda_A$.
The initial flux can than be obtained from the average path lengths in the $[0^-]$ and $[0^+]$ ensemble:
\begin{align}
f_A= \Big( \lo t_{\rm path}^{[0^-]} \rc
+ \lo t_{\rm path}^{[0^+]} \rc \Big)^{-1},
\end{align}
where $\lo t_{\rm path}^{[0^-]} \rc, \lo t_{\rm path}^{[0^+]} \rc$
are the average path
lengths in these ensembles. 
In this series of path ensembles, $\{ [0^-],[0^+],[1^+],\ldots,[(n-1)^+] \}$, 
the swapping move becomes 
extremely effective.
Note that the swapping moves
do not require any force calculations
except for the swapping between $[0^-]$ and  $[0^+]$ (see Fig.~\ref{swap}). Here, the last timestep
of the old path in
the $[0^-]$ ensemble is used as initial point to generate a new trajectory
in $[0^+]$ by integrating the equation of motion forward in time. Conversely,
the initial point of the old path in $[0^+]$ is followed backward in time
to generate a path in $[0^-]$. 
The TIS algorithm is then as follows. At each step it is decided
by an equal probability whether a series of shooting or swapping moves
will be performed. In the first case, all simulations will be updated
sequentially by a shooting move. In the second case, again an equal
probability will decide whether the swaps  $[0^-] \leftrightarrow [0^+],
[1^+] \leftrightarrow [2^+], \ldots$ or the swaps $[1^+] \leftrightarrow [2^+],
[3^+] \leftrightarrow [4^+], \ldots$ are performed~\cite{vanErp07PRL}.  

The effectiveness of this algorithm was illustrated in~\cite{vanErp07PRL} for the denaturation transition of DNA using 
the mesoscopic Peyrard-Bishop-Dauxois (PBD) model~\cite{PBD}. 
In this model, the DNA molecule is represented by a sequence of one dimensional particles representing the relative base-pair separations from the groundstate positions.  
Each particle is positioned in an external Morse potential describing the interaction of base-pairs of opposite strands. In addition, a first-neighbor anharmonic spring potential is used for the stacking interaction between bases of the same strand.  The width and depth of the Morse potential
are adapted to describe the weak AT or the strong GC interaction. Due to thermal fluctuations 
the hydrogen bonds between base-pairs of opposite strand can break, which
corresponds  to a particle moving on the plateau of the Morse potential. However, if the neighboring particles are still in the closed state, this particle will be rapidly pulled back into the 
stack. The fully denaturated state is achieved when all  base-pairs move on 
Morse plateau after 
which the two DNA strands have no interaction anymore and can move to infinite distances. 
This event is very rare for the larger molecules and has a very complex  dynamics  as it can proceed via different path ways. The DNA molecule might 
initially open up
at one end and propagate the opening through the molecule. 
Alternatively, a bubble in the middle might appear that continues to  
grow in both directions.         
Henceforth, the denaturation process is a typical example of a multiple 
reaction channel system and an accurate evaluation of the rate is quite a 
challenge for the larger molecules. 

In Ref.~[\onlinecite{vanErp07PRL}] the dynamics of a 20 base-pair DNA molecule of  AT bases was investigated by TIS with and without swapping. As RC, the base-pair separation of the base-pair
with the smallest distance  was used.  In total eight interfaces $\{\lambda_0, \ldots, \lambda_7 \}$ 
were defined (for more details see Ref.~[\onlinecite{vanErp07PRL}]).  
The calculated rates rates were in good agreement $0.0492 \pm 0.0062$  and $0.0524 \pm 0.0025$ ns$^{-1}$ for standard TIS  and TIS with path swapping respectively. The latter has a significant lower error despite a much shorter simulation. A very accurate integration method~\cite{VanErpPRL} 
for quasi one-dimensional 
systems confirmed the path swapping result within a 0.6 \% uncertainty.

A very instructive approach to
quantify the efficiency of the individual simulations and the simulation in total is 
given in~\cite{van06} by the introduction of so-called  
efficiency times $\tau_{\rm eff}$.  These are defined as the number of force calculations
required to obtain a statistical error equal to 1 in a given simulation.  
For the TIS simulations $[0^+],[1^+],\ldots$ these can 
be expressed as~\cite{van06}:
\begin{align}
\tau_{\rm eff}^{[i^+]}=\frac{1-p_i}{p_i} \xi_i L_i {\mathcal N}_i .
\label{eff}
\end{align}
Here, $p_i={\mathcal P}_A(\lambda_{i+1}|\lambda_i)$ and $L_i=\lo t_{\rm path}^{[i+]} \rc/\Delta t$ with
$\Delta t$ the MD timestep. These  are in principle independent to the simulation method. $\xi_i$ is the ratio between the average cost of the
simulation cycle and $L_i$.  ${\mathcal N}_i$ is the effective correlation between trajectories.

\begin{figure}[t]
\begin{center}
\includegraphics[width=7cm,angle=0,keepaspectratio]{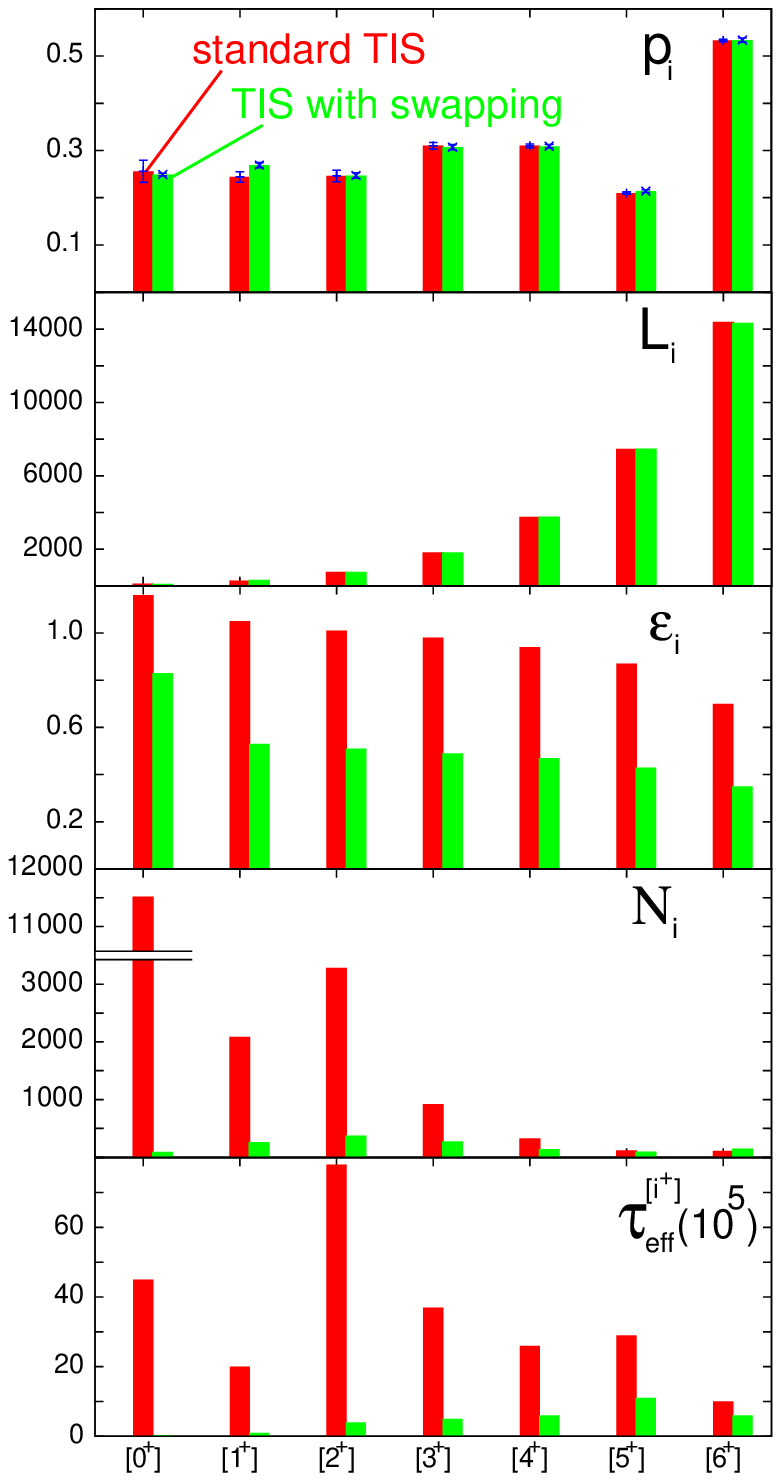}
\end{center}
\caption{(color online) Efficiency analysis for the TIS interface simulations with (green) and without (red) path swapping.
The crossing probability $p_i$, the average discrete path length $L_i$, the ratio $\xi_i$ and
the effective correlation ${\mathcal N}_i$ (note the change of scale in the y-axis) is shown in the first 4 panels. The last panel shows the effective computational cost per interface simulation. For the actual data see Ref.~[\onlinecite{vanErp07PRL}] }
\label{4fig}
\end{figure}

Fig.~\ref{4fig} shows the five parameters in Eq.~(\ref{eff}) for the seven TIS path ensembles
using standard TIS and path swapping.
The results for $p_i$ and $L_i$ are the same as expected.
 The values for $\xi_i$ show that the average cost per cycle is reduced 
 by a factor 1.5 for  the $[0^+]$ and by a factor 2 for the $[i^+], i>0$, simulations. 
This is due to the swapping moves that do not require any force calculations 
except for $[0^-] \leftrightarrow [0^+]$. 
However, more importantly is the dramatic reduction of  ${\mathcal N}_i$. The large values of 
${\mathcal N}_i$ in the standard TIS simulations are directly related 
to the problems of ergodic sampling when the system gets stuck for a long time in one specific 
reaction channel. Most spectacular is the reduction in the $[0^+]$ ensemble by more than two orders of magnitude. The reduction in both $\xi_i$ and ${\mathcal N}_i$ is reflected in $\tau_{\rm eff}^{[i^+]}$.
The computational cost is reduced at all levels by at least a factor of two, but
much more for ensembles $[0^+], \ldots,  [4^+]$ that consists of shorter paths with lower energy.
When all results are taken together, it can be concluded that 
the path swapping technique resulted in a gain of 
efficiency by more than a factor 20.  

\section{Conclusions} \label{seccon}
We have discussed the ability of path sampling to study transitions that proceed via multiple reaction channels. This is a common phenomenon  for any complex reaction mechanism that involves many degrees of freedom.  The analysis of these processes  are a huge challenge from a computational point of view. Besides all the difficulties 
of rare event simulations,  many additional issues come into play. 
The development of a feasible RC, that can be used in a standard free energy based method,
is notoriously difficult for high dimensional complex systems. Finding a RC that can well describe several reaction mechanisms simultaneously is even more difficult than that. 

Alternatively, one could rely 
on a method for which the correct RC does not play such a crucial role.    We have given several justifications 
why this is the case for the TIS and TPS path sampling methods. For a 2D model system of a slanted
barrier, the advantageous scaling of TIS compared to standard methods 
was proven rigorously when the effective computational cost 
as function of the 'quality of the RC'  was calculated analytically~\cite{van06}. In sec.~\ref{secRC} we gave some additional arguments based on a fictitious 
two channel problem displayed by the path-survival diagram
of Fig.~\ref{graph}.  
The difficulty of this example is that one channel is initially favorable but finally turns into a dead end.
If we assume a perfect ergodic sampling, the TIS simulations are still able to return the right result
using a minimal amount of trajectories. FFS, in which trajectories are propagated only forward in time,
would not be able to achieve the same. It can not change the history of the paths and will therefore
get stuck in  the unfavorable channel.
Apart from this, to be able to sample between 
several distinct reaction mechanisms, one needs to combine nonlocal MC moves with a high rate of acceptance. The shooting move has shown to have the required nonlocal characteristics~\cite{Geissler99}, but in order to sample efficiently,
transitions between the different channels need to occur at a much higher frequency than in the standard shooting algorithm. 

Sec.~\ref{secPS} shows a very promising approach to accomplish this based on the ideas
of parallel tempering~\cite{marinari92}. Instead of performing several path sampling simulations at different temperatures~\cite{VS01}, the swapping occurs between the different interface ensembles.
The initial MD simulation to compute the flux is replaced by an 'internal' path ensemble
which allows to swap trajectories at all levels in the system.
Compared to standard parallel tempering,
this gives the advantage that there is no need to extend the number of TIS simulations.  It also provides a way to overcome entropic barriers for which parallel tempering would not help. We reviewed the results of Ref.~[\onlinecite{vanErp07PRL}] where this method was applied on the DNA denaturation transition using the mesoscopic PBD~\cite{PBD} model.
These results showed that the PPS technique improved the TIS efficiency by more than a factor 20. 

Still, there are many issues that need to be studied  and more improvements might be possible. 
An exact estimate of the total error becomes more complicated as the standard error
propagation rules assume that the different simulations are independent. We think that
the presence of covariant terms will not have a huge effect, but this will be investigated explicitly 
in the near future.
In~\cite{van06} we derived relations for how to divide a total fixed simulation time
over a simulation series to obtain the lowest possible overall statistical error. The optimum was found 
when each simulation was given a simulation time proportional to $\propto \sqrt{\tau_{\rm eff}^{[i^+]}}$.
This result is generic for any type of method for which the final quantity is obtained by a product
outcomes from  a series of independent simulations. 
Finding the optimal way to divide the total simulation time in a PPS algorithm
is much more complicated. The simulations are no longer independent and each simulation has  a double function. Besides the evaluation of the intrinsic property, their function is also to assist the ergodic sampling of the other simulations.
So far, the PPS simulations were performed using the same number of trajectories
for  each ensemble. We are now developing  an improved PPS-TIS method that iteratively adapts the number of trajectories per ensemble to the optimal 
ratios~\cite{inprogress}. 
We think that TIS in combination with PPS can become an important standard 
method for the sampling of rare events in complex systems and multiple reaction channels in particular.

\section{Acknowledgements}
I would like to thank Pierre Mignon for carefully reading this paper.

\bibliographystyle{prsty}

\end{document}